\documentclass[reprint,prl,twocolumn,superscriptaddress,showpacs,showkeys,floatfix,preprintnumbers]{revtex4}

\pdfoutput=1

\usepackage{graphicx}

\usepackage{amssymb}
\usepackage{epsfig}
\usepackage{amssymb}
\usepackage{epsf}
\usepackage{epsfig}
\usepackage{color}
\usepackage{ifpdf}
\usepackage{longtable}

\newcommand{\nn}{\nonumber}
\newcommand{\ba}{\begin{eqnarray}}
\newcommand{\ea}{\end{eqnarray}}
\newcommand{\be}{\begin{equation}}
\newcommand{\ee}{\end{equation}}
\newcommand{\bd}{\begin{displaymath}}
\newcommand{\ed}{\end{displaymath}}

\newcommand{\chipt}{$\chi$PT}

\newcommand{\old}[1]{}

\newcommand{\plotangle}{0}

\newcommand{\plotsize}{0.40\textwidth}
\newcommand{\plotgap}{0.04\textwidth}

\begin{document}

\preprint{KEK-CP-271}
\preprint{OU-HET-743-2012}

\title{Two-photon decay of the neutral pion in lattice QCD}

\date{\today}

\author{Xu Feng}
\affiliation{High Energy Accelerator Research Organization (KEK), Tsukuba 305-0801, Japan}

\author{Sinya Aoki}
\affiliation{Graduate School of Pure and Applied Sciences, University of Tsukuba, Tsukuba 305-8571, Japan}

\author{Hidenori Fukaya}
\affiliation{Department of Physics, Osaka University, Toyonaka, Osaka 560-0043, Japan}

\author{Shoji Hashimoto}
\affiliation{High Energy Accelerator Research Organization (KEK), Tsukuba 305-0801, Japan}
\affiliation{School of High Energy Accelerator Science, The Graduate University for Advanced Studies (Sokendai), Tsukuba 305-0801, Japan}

\author{Takashi Kaneko}
\affiliation{High Energy Accelerator Research Organization (KEK), Tsukuba 305-0801, Japan}
\affiliation{School of High Energy Accelerator Science, The Graduate University for Advanced Studies (Sokendai), Tsukuba 305-0801, Japan}

\author{Jun-ichi Noaki}
\affiliation{High Energy Accelerator Research Organization (KEK), Tsukuba 305-0801, Japan}

\author{Eigo Shintani}
\affiliation{RIKEN-BNL Research Center, Brookhaven National Laboratory, Upton, NY 11973, USA}

\collaboration{JLQCD Collaboration}
\noaffiliation

\begin{abstract}
We perform non-perturbative calculation of the $\pi^0\to\gamma\gamma$
transition form factor and the associated decay width using lattice
QCD. The amplitude for two-photon final state, which is not an
eigenstate of QCD, is extracted through an Euclidean time integral of
the relevant three-point function. We utilize the all-to-all quark
propagator technique to carry out this integral as well as to include 
the disconnected quark diagram contributions. The overlap fermion
formulation is employed on the lattice to ensure exact chiral
symmetry on the lattice. After examining various sources of systematic
effects except for possible discretization effect, we obtain 
$\Gamma_{\pi^0\to\gamma\gamma}=7.83(31)(49)$~eV for the pion decay width,
where the first error is statistical and the second is our
estimate of the systematic error.
\end{abstract}



\maketitle

The neutral pion decay process provides a unique opportunity to test a
fundamental symmetry property of the gauge theory.
A quantum effect due to a fermion loop violates the axial-current
conservation, and gives the dominant contribution to the 
$\pi^0\to\gamma\gamma$ decay rate.
The prediction from this Adler-Bell-Jackiw (ABJ) anomaly~\cite{Adler:1969gk,Bell:1969ts} (or the
chiral anomaly) is rather precise because higher-loop diagrams do
not contribute in the limit of vanishing quark mass and external
momenta~\cite{Adler:1969er} even under the presence of strong interaction.
On the other hand, a recent experimental measurement of the neutral pion decay width
has reached the accuracy of 2.8\%~\cite{Larin:2010kq} and is aiming to achieve 1.4\%
in the near future.
At this level of accuracy the correction due to finite quark mass
becomes relevant.
Phenomenologically, an enhancement of the decay width of about 3--5\% has
been expected according to the sum rule and chiral perturbation theory ({\chipt})
approaches~\cite{Ioffe:2007eg,Goity:2002nn,Ananthanarayan:2002kj,Kampf:2009tk}, in which the main effect comes from a
mixing of $\pi^0$ with $\eta$ and $\eta'$ mesons. For a recent review, we refer the 
reader to Ref.~\cite{Bernstein:2011bx}.

In this letter we present a model-independent calculation of the
$\pi^0\to\gamma\gamma$ amplitude using the lattice Quantum
Chromodynamics (LQCD) including dynamical up, down and strange quarks.
We use the overlap fermion formulation~\cite{Neuberger:1997fp}, which 
preserves chiral symmetry at finite lattice spacings. In this formulation,
the chiral anomaly appears through the Jacobian of chiral transformation;
Atiyah-Singer's index theorem is reproduced as far as the background
gauge field is smooth enough~\cite{Fujikawa:2000qw}. 
On the other hand, whether the chiral anomaly is correctly reproduced 
at practically used lattice spacings ($\sim$ 0.1~fm) is a non-perturbative
problem, that we address in this work.

Compared to previous attempts~\cite{Cohen:2008ue,Shintani:2011vc}, 
a new technique is applied to treat two-external-photon state,
by utilizing the all-to-all quark propagator~\cite{Foley:2005ac,Kaneko:2010ru}.
In the limit of degenerate up and down quark masses we obtain the
decay rate with a statistical error of 4\% and a total error of 7\% 
after examining possible systematic 
effects.

The $\pi^0\rightarrow\gamma\gamma$ decay rate 
at the leading order of QED can be 
expressed as
\ba
\label{eq:width}
\Gamma_{\pi^0\gamma\gamma}=\frac{\pi\alpha_e^2m_\pi^3}{4}{\mathcal F}^2_{\pi^0\gamma\gamma}(m_\pi^2,0,0)\;,
\ea
where $\alpha_e$ is the fine structure constant, $m_\pi$ is the neutral pion mass and ${\mathcal F}_{\pi^0\gamma\gamma}(m_\pi^2,p_1^2,p_2^2)$ is the form factor of the pion to two (virtual) photon transition with $p_{1,2}$ the photon momenta.
In the chiral limit, 
the ABJ anomaly predicts
\ba
\label{eq:abj}
{\mathcal F}_{\pi^0\gamma\gamma}^{\rm ABJ}\equiv{\mathcal F}_{\pi^0\gamma\gamma}(0,0,0)=\frac{1}{4\pi^2F_0}\;,
\ea
where $F_0$ is the pion decay constant $F_\pi$ in the chiral limit. 
We define a normalized form factor as
$F(m_\pi^2,p_1^2,p_2^2)\equiv(4\pi^2F_\pi){\mathcal F}_{\pi^0\gamma\gamma}(m_\pi^2,p_1^2,p_2^2)$. In the Minkowski space-time ${\mathcal F}_{\pi^0\gamma\gamma}(m_\pi^2,p_1^2,p_2^2)$
is defined through the matrix element
\ba
\label{eq:def_Minkowski}
M_{\mu\nu}(p_1,p_2)&=&i\int d^4x\;e^{ip_1x}\langle\Omega|T\{j_\mu(x)j_\nu(0)\}|\pi^0(q)\rangle\nn\\
&=&\varepsilon_{\mu\nu\alpha\beta}p_1^\alpha p_2^\beta {\mathcal F}_{\pi^0\gamma\gamma}(m_\pi^2,p_1^2,p_2^2)
\ea
where $q$ is the $\pi^0$ momentum satisfying the on-shell condition $q^2=m_\pi^2$. The current $j_\mu=\sum_f Q_f \bar{q}_f\gamma_\mu q_f$ is the hadronic
component of the electromagnetic vector current and the sum runs over
all relevant quark flavors: $f=u,d,s$. $Q_f$ denotes the electromagnetic charge of them: $Q_u=+2/3$ and $Q_{d,s}=-1/3$. The factor $\varepsilon_{\mu\nu\alpha\beta}p_1^\alpha p_2^\beta$ is induced
by the negative parity of $\pi^0$.

By an analytic continuation of
(\ref{eq:def_Minkowski}) from the Minkowski to Euclidean space-time~\cite{Ji:2001wha,Dudek:2006ut}
one may write
\begin{widetext}
\ba
\label{eq:correlator}
&&M_{\mu\nu}(p_1,p_2)=\lim_{t_{1,2}-t_\pi\rightarrow\infty}\frac{1}{\frac{\phi_{\pi,\vec{q}}}{2E_{\pi,\vec{q}}}e^{-E_{\pi,\vec{q}}(t_2-t_\pi)}}
\int dt_1\;e^{\omega (t_1-t_2)}C_{\mu\nu}(t_1,t_2,t_\pi)\;,\nn\\
&&C_{\mu\nu}(t_1,t_2,t_\pi)\equiv\int d^3\vec{x}\;e^{-i\vec{p}_1\cdot\vec{x}}
\int d^3\vec{z}\;e^{i\vec{q}\cdot\vec{z}}
\langle\Omega|T\{j_\mu(\vec{x},t_1)j_\nu(\vec{0},t_2)\pi^0(\vec{z},t_\pi)\}|\Omega\rangle\;,
\ea
\end{widetext}
where $t_1$, $t_2$ and $t_\pi$ are Euclidean time slices. $\int {d^3\vec{z}}\;e^{i\vec{q}\cdot\vec{z}}\pi^0(\vec{z},t_\pi)$ is an interpolating operator for the neutral pion with the spatial momentum $\vec{q}$.
Its amplitude and energy in the ground state are denoted by $\phi_{\pi,\vec{q}}$ and
$E_{\pi,\vec{q}}$, respectively. The four-momentum of the first photon $p_1=(\omega,\vec{p}_1)$
is chosen as input, while the momentum of the second photon is given
 as $p_2=(E_{\pi,\vec{q}}-\omega,\vec{q}-\vec{p}_1)$ by momentum conservation.
Note that the analytical continuation is valid only for $p_{1,2}^2<M_V^2$, which sets the limit on the value of $\omega$.
($M_V$ stands for the invariant mass of the lowest energy state in the vector channel.) Since the two photons can not be on-shell simultaneously, we calculate the form factor at the off-shell photon momenta and then extrapolate to the on-shell limit.

To calculate the matrix element 
$\langle\Omega|j_\mu j_\nu \pi^0 |\Omega\rangle$ in Eq.~(\ref{eq:correlator}), 
we use 2+1-flavor
overlap fermion configurations generated by
the JLQCD and TWQCD Collaborations~\cite{Matsufuru:2008zz,Noaki:2010zz} 
at a single lattice spacing $a=0.11$ fm and 
two spatial lattice sizes $L/a=16$ and 24. The time extent is $T/a=48$.
Although the main gauge ensembles have a fixed (global) topological charge
$Q=0$, the deviation from the $\theta$-vacuum is
understood as a finite volume effect of $O(1/L^3T)$~\cite{Aoki:2007ka}. 
We check the significance of this effect by comparing the results with two different values $Q=0,1$.
We utilize the all-to-all propagator to
calculate the correlation function $C_{\mu\nu}(t_1,t_2,t_\pi)$ at
any time slices of $t_1$, $t_2$ and $t_\pi$.
 The electromagnetic current $j_\mu=\sum_fQ_f\bar{q}_f\gamma_\mu q_f$ is implemented on the lattice as
a local operator with a renormalization factor calculated nonperturbatively in~\cite{Noaki:2009xi}
to match the lattice results with the continuum theory.
The up and down quarks 
are degenerate in mass. We 
use the bare values $am_{u,d}=0.015$, $0.025$, $0.035$ and $0.050$, corresponding to
the pion mass $m_\pi$ ranging from
290 to 540 MeV. 
Our final results are obtained by an extrapolation of the data to the the physical pion mass $m_{\pi,\rm phy}$.
The strange quark mass
is fixed at $am_s=0.080$, which is very close to the estimated physical value.

\begin{figure}
\includegraphics[width=\plotsize,angle=\plotangle]{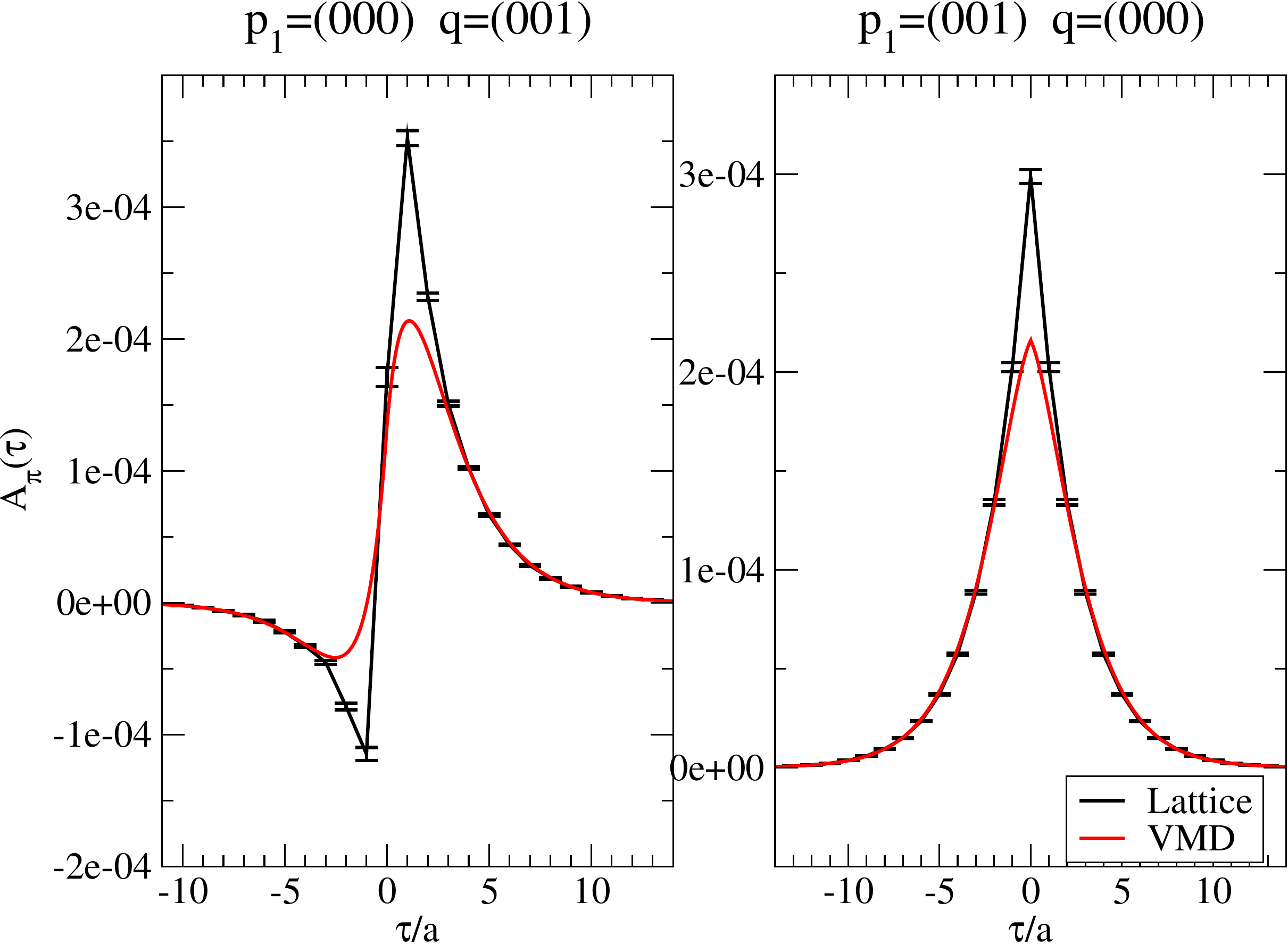}\hspace{\plotgap}
\caption{The amplitude $A_{\pi}(\tau)$ as a function of $\tau$ for momentum setups \#1 (left) and \#2 (right).
The black (red) curves indicate the lattice (VMD) amplitudes.}
\label{fig:distribution}
\end{figure}
From the large $t_{1,2}-t_\pi$ behavior of $C_{\mu\nu}(t_1,t_2,t_\pi)$,  
it is possible to extract the $\pi^0$-ground state.
We define the amplitude $A_\pi$ as
\ba
A_\pi(\tau)\equiv\lim_{t-t_\pi\rightarrow\infty}C_{\mu\nu}(t_1,t_2,t_\pi)/e^{-E_{\pi,\vec{q}}(t-t_\pi)}\;,
\ea
with $\tau=t_1-t_2$ and $t=\textmd{min}\{t_1,t_2\}$, and obtain $M_{\mu\nu}(p_1,p_2)$ by performing an integral
\ba
\label{eq:R_expression}
&&\hspace{-0.6cm}\frac{2E_{\pi,\vec{q}}}{\phi_\pi}\left(\int_0^{\infty} d\tau\;e^{\omega \tau}A_{\pi}(\tau)
+\int_{-\infty}^0 d\tau\;e^{(\omega-E_{\pi,\vec{q}}) \tau}A_{\pi}(\tau)\right)\;.\nn\\
\ea
We use two momentum setups
$\vec{p}_1=\frac{2\pi}{L}(0,0,0)$, $\vec{q}=\frac{2\pi}{L}(0,0,1)$ (setup \#1) and 
$\vec{p}_1=\frac{2\pi}{L}(0,0,1)$, $\vec{q}=\frac{2\pi}{L}(0,0,0)$ (setup \#2). The resulting amplitudes $A_{\pi}(\tau)$ for these setups
are shown in Fig.~\ref{fig:distribution}.
In order to qualitatively understand the $\tau$-dependence of the $A_{\pi}(\tau)$,
we consider the
vector-meson-dominance (VMD) model
${\mathcal F}_{\pi^0\gamma\gamma}^{\rm VMD}(m_\pi^2,p_1^2,p_2^2)=c_VG_V(p_1^2)G_V(p_2^2)$,
with $G_V(p^2)=M_V^2/(M_V^2-p^2)$ the vector meson propagator and $c_V$ a constant.
The amplitude $A_{\pi}^{\rm VMD}(\tau)$, reconstructed from this model, is 
plotted by red curves in Fig.~\ref{fig:distribution}. (The detailed expression for $A_\pi^{\rm VMD}(\tau)$ will be given in a later publication~\cite{Feng:long_paper}.) 
We find that the VMD model 
describes the lattice data already at $|\tau|/a=7$, and we can
safely evaluate the contribution beyond $|\tau|/a=13$, where the lattice data are
truncated due to the finite time extent $T$.
At small $|\tau|$ the VMD model fails to match the lattice data.
This is because no information of the vector-meson excited states
are contained in ${\mathcal F}_{\pi^0\gamma\gamma}^{\rm VMD}(m_\pi^2,p_1^2,p_2^2)$.
Given the dominant role played by the lowest vector meson,
we take its form as a basis to analyze the functional form of ${\mathcal F}_{\pi^0\gamma\gamma}(m_\pi^2,p_1^2,p_2^2)$.
Namely, we use an expression
\ba
\label{eq:expansion}
&&{\mathcal F}_{\pi^0\gamma\gamma}(m_\pi^2,p_1^2,p_2^2)=c_VG_V(p_1^2)G_V(p_2^2)\nn\\
&&\hspace{1cm}+\sum_{m} c_m
\left((p_2^2)^mG_V(p_1^2)+(p_1^2)^mG_{V}(p_2^2)\right)\nn\\
&&\hspace{1cm}+\sum_{m,n}c_{m,n}(p_1^2)^m(p_2^2)^n\;,
\ea
which includes possible contributions from excited states as a polynomial of $p_{1,2}^2$.
In the chiral and large volume limit
for which the two-pion threshold opens, the VMD model would no longer
give an adequate description of the data. Such effect could become
significant for precision better than few percent, which is not within the
scope of this work.

\begin{figure}
\includegraphics[width=\plotsize,angle=\plotangle]{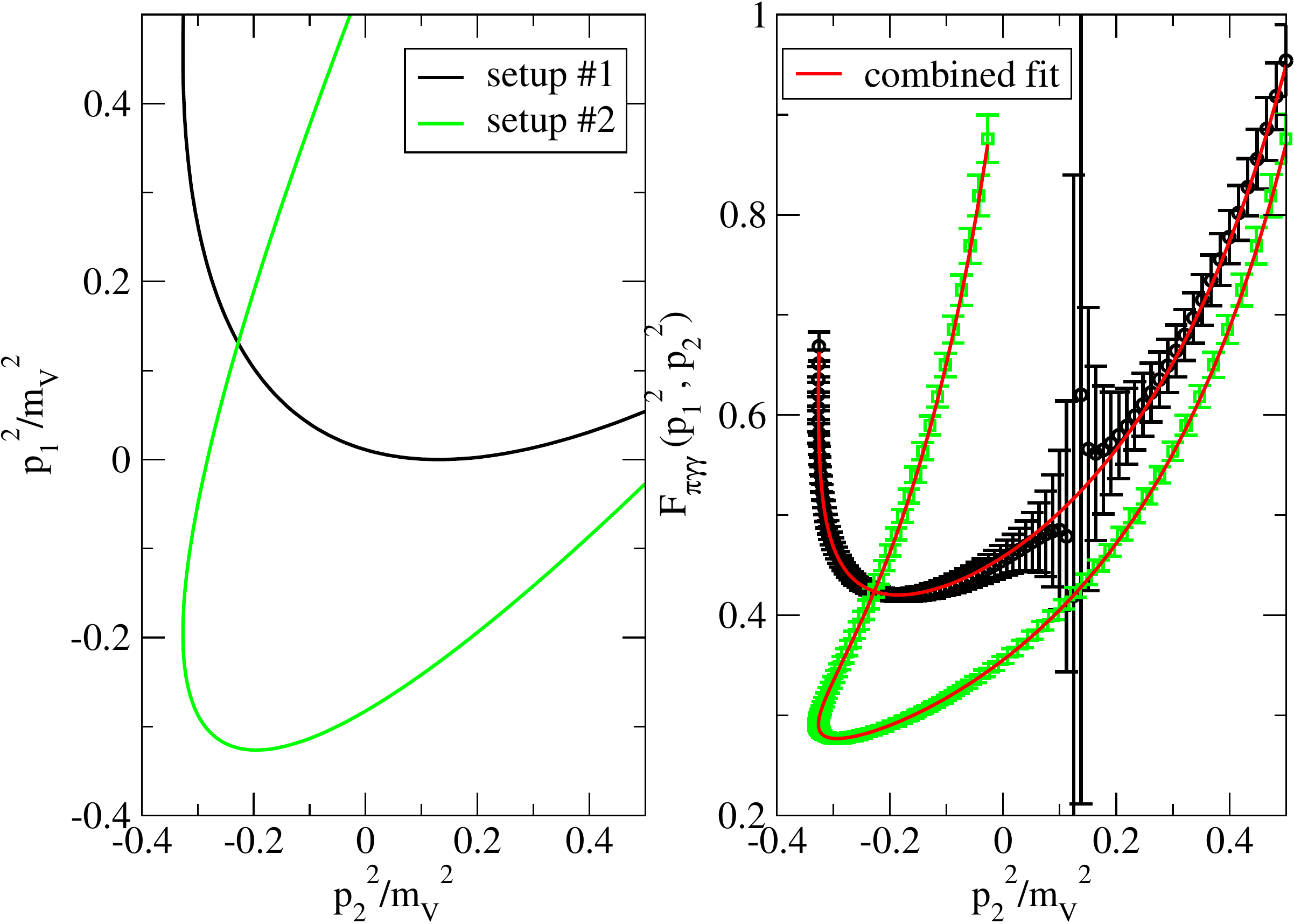}\hspace{\plotgap}
\caption{Left: Contour of $(p_1^2,p_2^2)$ rescaled by $1/M_V^2$ for our momentum setups.
Right: $F_{\pi^0\gamma\gamma}(m_\pi^2,p_1^2,p_2^2)$
as a function of $p_2^2/M_V^2$. Lattice data are obtained on a $24^3\times48$ lattice at $am_{u,d}=0.015$.}
\label{fig:matrix_element}
\end{figure}
\begin{figure}
\includegraphics[width=\plotsize,angle=\plotangle]{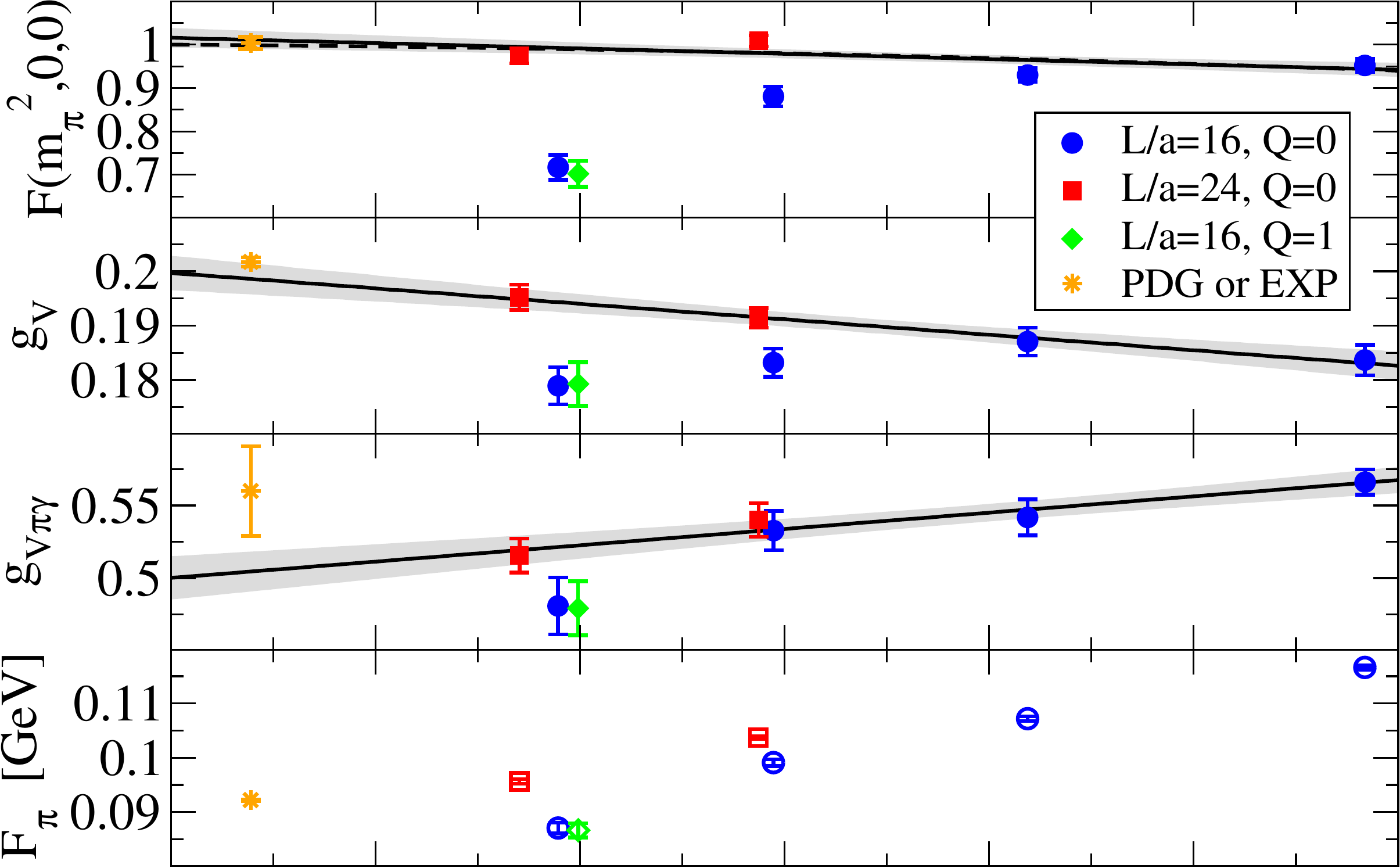}
\includegraphics[width=\plotsize,angle=\plotangle]{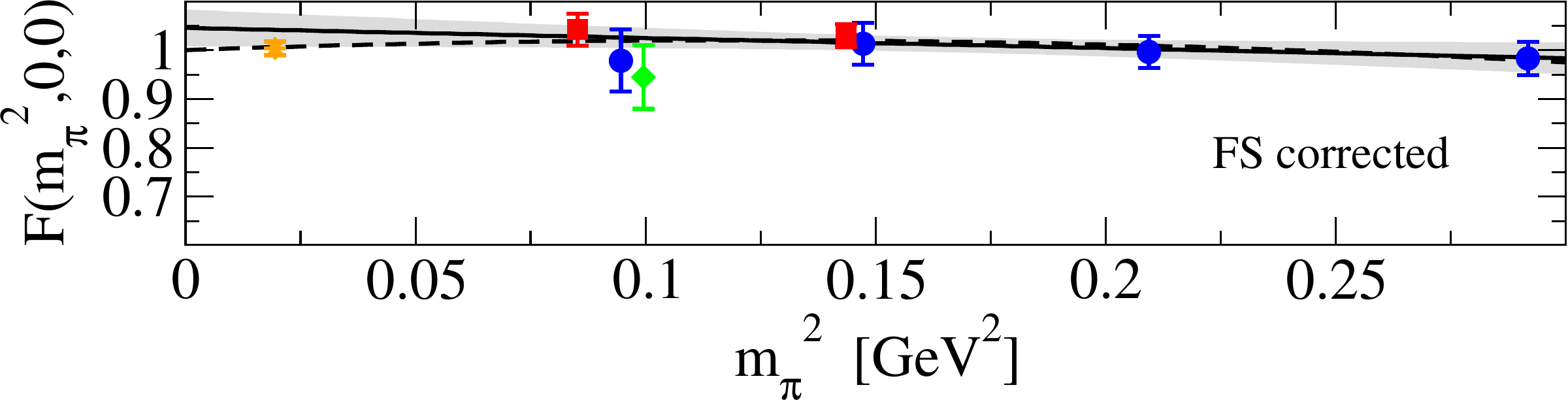}
\caption{$F(m_\pi^2,0,0)$, $g_V$, $g_{V\pi\gamma}$, $F_\pi$ and FS corrected $F(m_\pi^2,0,0)$ as a function of $m_\pi^2$ from top to bottom panels. In each panel, data with $(L/a,Q)=(16,0)$, $(24,0)$ and $(16,1)$ are plotted by the blue, red and green symbols, respectively. The yellow symbols indicate the Particle Data Group (PDG)~\cite{Nakamura:2010zzi} or PrimEx~\cite{Larin:2010kq} experimental values for reference.
The solid (dashed) curves show the result of the fit to the linear (quadratic) function.
The dataset used in the fit is explained in the text.}
\label{fig:ff_wo_FS}
\end{figure}

By varying $\omega$, we obtain $M_{\mu\nu}(p_1,p_2)$ in a certain range of $p_1^2$ and $p_2^2$.
As shown in the left panel of Fig.~\ref{fig:matrix_element}, a pair $(p_1^2,p_2^2)=(\omega^2-\vec{p}_1^2,
(E_{\pi,\vec{q}}-\omega)^2-(\vec{q}-\vec{p}_1)^2)$ forms a continuous contour on
the $(p_1^2,p_2^2)$ plane for $p_{1,2}^2<M_V^2/2$.
Evaluating ${\mathcal F}_{\pi^0\gamma\gamma}(m_\pi^2,p_1^2,p_2^2)$ along this contour,
we obtain the data plotted in the right panel of Fig.~\ref{fig:matrix_element}.
We perform the combined fit of these data to Eq.~(\ref{eq:expansion}) with four free parameters: $c_V$, $c_0$,
$c_{0,0}$ and $c_{0,1}=c_{1,0}$, truncating the higher-order terms which turned out to be 
negligibly small.
The fitting curves are shown in the right panel
of Fig.~\ref{fig:matrix_element}.
As expected, the single formula~(\ref{eq:expansion}) describes the data with different 
momentum setups.
Combining the resulting fit parameters, we obtain 
the normalized form factors $F(m_\pi^2,0,0)$, which are plotted in the uppermost panel of 
Fig.~\ref{fig:ff_wo_FS}.

In the following we analyze the details of systematic effects.
When calculating the integral in Eq.~(\ref{eq:R_expression}),
we use the summation 
instead of the integration. 
This causes a discretization effect, which vanishes in the continuum limit.
Putting $A_\pi^{\rm VMD}(\tau)$ into Eq.~(\ref{eq:R_expression}), 
we find that the fractional difference between $M_{\mu\nu}(p_1,p_2)$ from the summation 
and the integration is less than $5\times10^{-4}$. 
With the lattice data that include the excited state contributions, we could expect a larger error,
$\sim1\times10^{-3}$, which is estimated from a difference between VMD and lattice data in Fig.~\ref{fig:distribution}.
We can therefore safely neglect this source of error as it is
well below 1\%. 

We use two lattice volumes and two topological-charge sectors to check 
finite-size (FS) effects. 
Following Ref.~\cite{Aoki:2007ka} we analyze the fixed-topology (FT) effect
and find it suppressed due to the kinematical structure of 
$\varepsilon_{\mu\nu\alpha\beta}p_1^\alpha p_2^\beta$.
By comparing the lattice results at different topological-charge sectors,
we do not observe statistically significant FT effects.
The leading FS effect in $C_{\mu\nu}(t_1,t_2,t_\pi)$ is the conventional one and known to behave as $e^{-m_\pi L}$~\cite{Gasser:1987zq}. 
To reduce the contamination due to this effect,
we therefore use the data with $m_\pi L \ge 4$ 
to perform the chiral extrapolation. (Namely, we exclude the $L/a=16$ data points at the lowest two pion masses.)

{\chipt} shows that up to next-to-leading order (NLO)
the $m_\pi$-dependence of $F(m_\pi^2,0,0)$ involves no chiral logarithm~\cite{Donoghue:1986wv,Bijnens:1988kx}.
We therefore simply fit $F(m_\pi^2,0,0)$ by a linear function in $m_\pi^2$, and obtain
$F(0,0,0)=1.016(20)$ and $F(m_{\pi,\rm phy}^2,0,0) = 1.011(19)$. 
To check the higher-order correction, we also perform a quadratic fit
under the constraint from the ABJ anomaly: $F(0,0,0)=1$. 
We do not find any statistically significant difference due to the higher-order term. The linear (quadratic) fit is shown by the solid (dashed) 
line in the uppermost panel of Fig.~\ref{fig:ff_wo_FS}.

Next we consider the data with $m_\pi L<4$, which tend to suffer from the FS effect.
As shown in Fig.~\ref{fig:ff_wo_FS}, at $m_\pi\approx 290$ MeV we find that $F(m_\pi^2,0,0)$
 calculated at $L/a = 16$ lattice is 27\% less than the one at $L/a = 24$.
Although large, such FS effect is understandable.
By inserting the ground state into $\langle j_\mu j_\nu \pi^0\rangle$, 
we can approximate this three-point correlation function with three hadronic matrix elements:
\bd
\langle j_\mu j_\nu \pi^0\rangle \rightarrow \langle \Omega|j_\mu|V,\varepsilon\rangle\langle V,\varepsilon |j_\nu |\pi^0\rangle\langle\pi^0|\pi^0|\Omega\rangle\;.
\ed
The first matrix element is related to the electromagnetic coupling $g_V$ 
as $\langle\Omega|j_\mu|V,\varepsilon\rangle=M_V^2 g_V \varepsilon_\mu$, 
the second is proportional to the $V\pi\gamma$ coupling $g_{V\pi\gamma}$ 
and the third is related to $F_\pi$ by the PCAC relation. 
In our calculation we do not observe significant FS effect in $M_V$ but
find 8\%, 7\% and 9\% shifts in $g_V$, $g_{V\pi\gamma}$ and $F_\pi$, respectively, from $L/a=16$ to 24, as shown in Fig.~\ref{fig:ff_wo_FS}. 
These FS effects may accumulate in the three-point function.
We estimate the FS corrections 
$R_{{\mathcal O}}\equiv {\mathcal O}(\infty)/{\mathcal O}(L)$ with ${\mathcal O}=g_V$, $g_{V\pi\gamma}$ and $F_\pi$. 
$R_{g_V}$ and $R_{g_{V\pi\gamma}}$ are evaluated by adding a correction term, $e^{-m_\pi L}$, into the linear fit form in the chiral extrapolation of each quantity. With such corrections taken into account we confirm that their chiral limit is consistent with experimental data.
$R_{F_\pi}$ is calculated to NNLO by using {\chipt}~\cite{Colangelo:2005gd}.
Assuming that $R_{F(m_\pi^2,0,0)}=R_{g_V}R_{g_{V\pi\gamma}}R_{F_\pi}$ we may correct $F(m_\pi^2,0,0)$ by a factor of $R_{F(m_\pi^2,0,0)}$. 
As shown in the lowest panel of Fig.~\ref{fig:ff_wo_FS} with FS correction $F(m_\pi^2,0,0)$ at $L/a=16$ agrees with 
those at $L/a=24$. Using the corrected data to perform a linear extrapolation, we obtain
$F(0,0,0)=1.045(35)$ and $F(m_{\pi,\rm phy}^2,0,0)=1.041(32)$.
The difference between the results from the two methods is considered 
as a systematic error.

\begin{figure}
\includegraphics[width=\plotsize,angle=\plotangle]{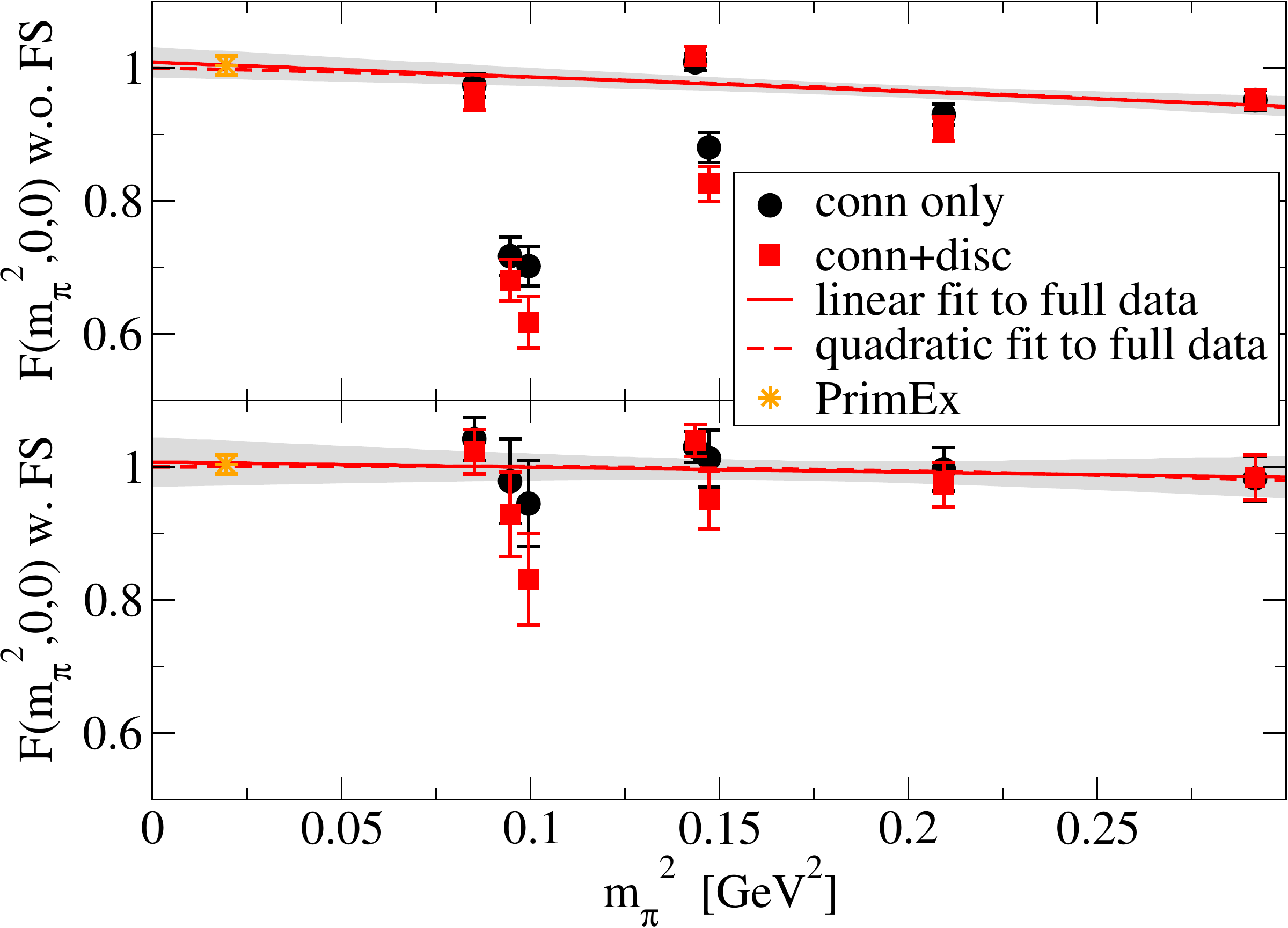}\hspace{\plotgap}
\caption{A comparison of the full contribution to $F(m_\pi^2,0,0)$ and the connected-only piece.
The upper (lower) panel shows the result without (with) FS correction.}
\label{fig:conn_vs_disc}
\end{figure}

So far, our results are obtained neglecting the effect of disconnected diagrams that may appear because the electromagnetic
current $j_\mu$ contains flavor-singlet contribution. Calculation of the disconnected diagram is computationally demanding and statistically noisy. 
We solve these problems by the use of the all-to-all propagator. 
The full data, including both the connected and disconnected contributions, are plotted in the upper (lower) panel of Fig.~\ref{fig:conn_vs_disc} for the case without (with) the FS correction. We find that, although not significant, there is a
shift from the connected data to the full ones. Since the accuracy of our calculation
reaches a few-percent level, the disconnected effect is relevant. 
Using the full data we repeat the analysis.
The linear fit of $F(m_\pi^2,0,0)$ with $m_\pi L\ge4$ yields
$F(0,0,0) = 1.009(22)$ and $F (m_{\pi,\rm phy}^2,0,0) = 1.005(20)$. The fit with
FS corrected $F(m_\pi^2,0,0)$ produces
$F (0,0,0) = 1.007(36)$ and  $F (m_{\pi,\rm phy}^2,0,0) = 1.006(33)$.
Including the disconnected contributions, 
the normalized form factor in the chiral limit and at the physical pion mass
shifts by 1-4\%.
This is comparable to the statistical error. 

Using the full data we quote our
results for $F(m_\pi^2,0,0)$ and $\Gamma_{\pi^0\gamma\gamma}$ in the isospin symmetric limit as
\ba
F (0,0,0)&=& 1.009(22)(29)\;,\nn\\
F (m_{\pi,\rm phy}^2,0,0) &=& 1.005(20)(30)\;,\nn\\
\Gamma_{\pi^0\gamma\gamma}&=&7.83(31)(49)\;\textmd{eV}\;.
\ea
where the systematic errors originate from the 
difference of the results by using two methods of treating the FS effect. 
(The difference appearing
in the full data is small. To be conservative, we use the connected  
data to estimate such systematic error.)
Our results reproduce the predication of the ABJ anomaly $F(0,0,0)=1$ and
agree with the PrimEx measurement $\Gamma_{\pi^0\gamma\gamma}=7.82(22)$ eV~\cite{Larin:2010kq}.
For future improvements, 
isospin breaking effects due to the light quark mass difference need to be included.

\begin{acknowledgments}
Numerical simulations are performed on the Hitachi SR16000 at Yukawa Institute of Theoretical Physics and
at High Energy Accelerator Research Organization
under a support of its Large Scale Simulation Program
(No. 10-11). This work is supported in part by the Grant-in-Aid of
the Japanese Ministry of Education (No.21674002, 21684013),
the Grant-in-Aid for Scientific Research on Innovative Areas
(No. 2004: 20105001, 20105002, 20105003, 20105005, 23105710), and the HPCI
Strategic Program of Ministry of Education.
\end{acknowledgments}

\bibliography{pi2gg_short}
\bibliographystyle{h-physrev}

\end{document}